\documentclass{bulsao}
\usepackage{graphicx}
\usepackage{latexsym}

\begin{document}

\title{Nearby low-mass triple system GJ\,795}

\author{E.~V. Malogolovets, Yu.~Yu. Balega, D.~A. Rastegaev}

\institute{Special Astrophysical Observatory, RAS, Nizhnii Arkhyz,
Karachai-Cherkessian Republic, 357147 Russia}

\offprints{E.~V. Malogolovets, \email{evmag@sao.ru}}

\date{received:September 21, 2006/revised: November 24, 2006}

\titlerunning{Nearby low-mass triple system GJ\,795}

\authorrunning{Malogolovets  et al.}

\abstract{
We report the results of our optical speckle interferometric observations of
the nearby triple system GJ\,795 performed with the 6 m BTA telescope  with
diffraction-limited angular resolution. The three
components of the system were optically resolved for the first time. Position measurements allowed us
to determine the elements of the inner orbit of the triple system. We use the
measured magnitude differences to estimate the absolute magnitudes and spectral
types of the components of the triple: $M_{V}^{Aa}$=7.31$\pm$0.08,
$M_{V}^{Ab}$=8.66$\pm$0.10, $M_{V}^{B}$=8.42$\pm$0.10, $Sp_{Aa}$ $\approx$K5,
$Sp_{Ab}$ $\approx$K9, $Sp_{B}$ $\approx$K8. The total mass of the system is
equal to $\Sigma\mathcal{M}_{AB}$=1.69$\pm$0.27$\mathcal{M}_{\odot}$. We show
 GJ\,795 to be a hierarchical triple system which satisfies the empirical
stability criteria.}

\maketitle

\section{INTRODUCTION}

According to the most recent concepts, stars form in small groups and clusters.
The disruption of such groups results in the formation of both multiple systems
and single stars. Stars in triple and more complex multiple systems make up for
more than  20$\%$ of the Milky-Way population. The study of their dynamical and
physical parameters is necessary for understanding the process of star
formation as a whole. However, the now available observational data are
insufficient for testing the theories of formation and evolution of multiple
stars. We do not yet entirely understand the initial conditions and mechanisms
of multiple star formation. The issues that still remain unclear include the
conservation of angular momentum in the process of star formation; the
dynamical stability of multiple systems with more than three stars; the effect
of tidal interactions on the dynamical evolution of multiple systems; the
distribution of orbital periods, eccentricities, component mass ratios, and
correlations between these parameters; mutual orientations of the orbital
planes in multiple systems, etc.

Of special interest is the study of multiple systems with low degree of
hierarchy with comparable orbital periods and semimajor axes. The Orion
Trapezium --- a small cluster of very young and massive stars
--- is the most well known dynamically unstable multiple system. The
disruption time scale of the Trapezium is estimated at $10^{4}-10^{6}$ years
(\cite{pflamm:Malogolovets_n}). Main-sequence stars are also found in a number of
systems that are potentially dynamically unstable
(\cite{fek:Malogolovets_n}; \cite{seb:Malogolovets_n}; \cite{tok1:Malogolovets_n}). The stability
criteria for multiple stellar systems were analyzed by
\cite{eggl:Malogolovets_n}; \cite{gol1:Malogolovets_n}; \cite{gol2:Malogolovets_n};
\cite{harr1:Malogolovets_n}; \cite{harr2:Malogolovets_n}; \cite{mardling:Malogolovets_n}
and other authors. However, the conclusion about the dynamical instability of a
particular system is often disproved when its orbital elements are refined. The
triple system ADS\,16904 with the periods of the inner and outer orbits equal
to 15 and 150 years, respectively, is an example (\cite{bal1:Malogolovets_n}).
According to all known criteria, the state of this system must be close to
instability (\cite{orlov:Malogolovets_n}). However, new interferometric
measurements including the data obtained with the 6 m telescope of the Special
Astrophysical Observatory of the Russian Academy of Sciences indicate that
the actual period of the outer orbit in  this triple is twice longer and hence
the system must be dynamically stable. So far, no systems with main-sequence
components have been found that could be securely classified as dynamically
unstable.

Known candidate objects with low orbit hierarchy, which therefore should be
viewed as possible dynamically unstable multiple systems, include the nearby
($d$ $\approx$16 pc) triple star GJ\,795 (HD\,196795 = Hip\,101955, $\alpha$ =
20$^{h}$39$^{m}$38$^{s}$, $\delta$ = +04$^{\circ}$58$'$19$''$, epoch 2000.0).
Its integrated spectral type corresponds to that of a  K5V star. For decades,
GJ\,795 has been known as the visual pair Kui\,99 (\cite{kui:Malogolovets_n})
with a period of 40 years. During his spectroscopic survey of visual binaries
with the CORAVEL radial-velocity scanner, Duquennoy (\cite{duq:Malogolovets_n})
found GJ\,795 to contain a hitherto unknown spectroscopic subsystem with very
low amplitude of radial-velocity variations. He concluded that the companion is
bound to the main component of the binary and computed a preliminary model of
the system by combining photometric and spectroscopic data with the computed
orbital elements. At the same time, the inclination of the inner orbit of the
triple  was estimated based on the assumed component masses exclusively. The
strongly inclined outer orbit also remained highly uncertain. To refine the
pattern of component motions in GJ\,795, this system was put in 1998 into the
list of program stars for speckle interferometric observations with the 6 m
telescope of the Special Astrophysical Observatory of the Russian Academy of
Sciences.

In this paper, we report the results of our speckle interferometric
observations of the relative measurements of components of  GJ\,795 and their
differential photometry, and determine the parameters of orbital motion of
stars and their dynamic masses. In conclusion, we discuss the dynamical
stability of the system.

\section{OBSERVATIONS AND DATA ANALYSIS}

\begin{figure}[tbp]
\begin{center}
\includegraphics[scale=0.6]{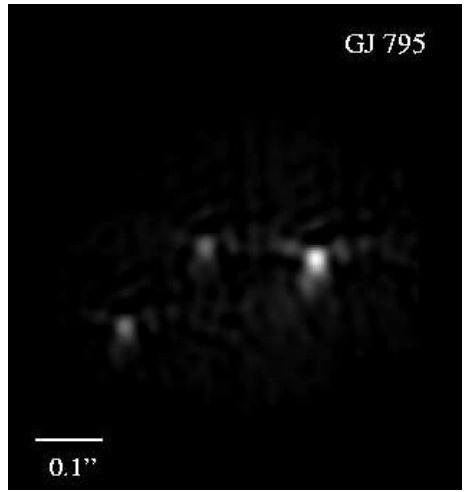}
\caption{The 610/20 nm image of GJ\,795 (1998.77)
reconstructed using bispectral analysis. Artefacts surrounding the point
sources are due to various types. North is at the top and East on the left.}
\label{fig1:Malogolovets_n}
\end{center}
\end{figure}

We performed speckle interferometric observations of
GJ\,795 with the new facility mounted on the 6 m telescope of the Special
Astrophysical Observatory of the Russian Academy of Sciences
(\cite{max:Malogolovets_n}). Its detector consists of a fast 1280$\times$1024
Sony ICX085 CCD combined with a three-camera image-tube converter with
electrostatic focusing. We recorded speckle interferograms in the visible part
of the spectrum with exposures ranging from  5 to 20 milliseconds.
Table~\ref{tab1:Malogolovets_n} lists the log of observations, which gives for
each measurement the date of observation (as a fraction of Besselian year);
seeing $\beta$ in arcsec; the number of speckle interferograms in the series;
filter parameter $\lambda/\Delta\lambda$ in nm, where $\lambda$ and
$\Delta\lambda$ are the central wavelength and half-bandwidth, respectively. We
determined the relative component positions and magnitude differences from the
power spectra of the speckle interferograms averaged over the series
\cite{bal2:Malogolovets_n}.
\begin{table}
\begin{center}
\caption{Log of speckle observations} \label{tab1:Malogolovets_n}
\bigskip
\begin{tabular}{c|c|c|c}
\hline
 Date      &  $\beta$    & N       & $\lambda/\Delta\lambda$ \\
           &  arcsec     &         & nm                      \\
\hline
1998.7741  &  1          & 700     & 610/20                  \\
1999.8206  &  2          & 1500    & 610/20                  \\
2000.8752  &  1.5        & 1000    & 600/30                  \\
2001.7522  &  2          & 1500    & 545/30                  \\
           &  2          & 1500    & 850/75                  \\
2002.7986  &  3-5        & 1500    & 600/30                  \\
2003.9272  &  1          & 2000    & 545/30                  \\
           &  1          & 2000    & 700/30                  \\
           &  1          & 2000    & 800/110                 \\
2004.8232  &  1.5        & 2000    & 600/30                  \\
\hline
\end{tabular}
\end{center}
\end{table}
The accuracy of the measured position parameters is 0.2--4.0$^{\circ}$ and 1--4
milliarcsec for position angle and angular separation, respectively.
Measurement errors depend on a number of parameters: component separation,
magnitude difference, and seeing $\beta$. The accuracy of the magnitude
differences inferred from reconstructed power spectra depends on the same
parameters. For objects in the  $m_{V}$=8--10 magnitude interval it usually
varies from 0.05 to 0.2. We use bispectral analysis of the interferogram series
to perform complete reconstruction --- including that of modulus and phase ---
of the images (\cite{lohm:Malogolovets_n}; \cite{weigelt:Malogolovets_n}).
Figure~\ref{fig1:Malogolovets_n} shows the reconstructed image of the triple
star GJ\,795 based on observations made in 1998.

\section{ABSOLUTE MAGNITUDES AND SPECTRAL TYPES}

As we already mentioned in the Introduction, Duquennoy
(\cite{duq:Malogolovets_n}) carried out a detailed study of the radial-velocity
variations of primary component A of the visual binary KUI\,99 with the CORAVEL
radial-velocity scanner. He detected no traces of the fainter star B in the
spectra. On two nights in 1985, significant variations were observed in the
profile of the correlation minimum that were due to the contribution of
component Ab to the total flux. The resulting radial-velocity curve was used to
determine the orbit of inner binary Aab with a period of $P$=920.2 days and
eccentricity $e$=0.747. The preliminary model of the system, which included all
three components, assumed a total mass and parallax of
$\Sigma\mathcal{M}_{AB}$=1.62$\pm$0.27$\mathcal{M}_{\odot}$ and $\pi$=64$\pm$5
mas, respectively. This model also made use of the empirical ``mass --
luminosity'' relation for  K6V (Aab) and K9V(B) type stars and a highly
uncertain visual orbit of the outer pair AB \cite{baize:Malogolovets_n}.

According to Hipparcos (\cite{eca:Malogolovets_n}) data, the trigonometric
parallax of  GJ\,795 differs significantly from the above value
($\pi_{Hip}$=53.82$\pm$2.21 mas). However, Hipparcos trigonometric parallaxes
for binary and multiple stars are known to be potentially fraught with extra
errors due to wrong correction of component orbital motions in the process of
Hipparcos data reduction (\cite{bal3:Malogolovets_n}; \cite{sha:Malogolovets_n}).
Soderhelm (\cite{soderh:Malogolovets_n}) corrected the parallax for the effect of
the orbital motion of the pair AB: $\pi_{Hip}^{\ast}$=58.8$\pm$2.1 mas. The
corrected Hipparcos parallax agrees within the quoted errors with that given by
Duquennoy (\cite{duq:Malogolovets_n}).

\begin{table}
\caption{Differential speckle interferometry of GJ\,795}
\label{tab2:Malogolovets_n}
\bigskip
\begin{tabular}{c|l|c|c|c|c}
\hline
Date      & Comp.     & $\Delta$m & $\sigma_{\Delta m}$ & $\lambda$/$\Delta\lambda$ & Reference \\
BY        & vector  &           &                     &   nm                      &        \\
\hline
1998.7741 & Aa-Ab      & 1.09      & 0.05                & 610/20                    &  Balega et al. \\
          & Aa-B       & 0.88      & 0.05                &                           &   2002a        \\
1999.8206 & Aa-Ab      & 1.14      & 0.03                & 610/20                    &  Balega et al. \\
          & Aa-B       & 0.94      & 0.03                &                           &    2004          \\
2000.8752 & Aa-Ab      & 1.30      & 0.06                & 600/30                    &   Balega et al.\\
          & Aa-B       & 1.02      & 0.06                &                           &     2006         \\
2001.7522 & Aa-Ab      & 1.35      & 0.06                & 545/30                    &  Balega et al. \\
          & Aa-B       & 1.11      & 0.06                &                           &       2006    \\
          & Aa-Ab      & 0.92      & 0.06                & 850/75                    &   Balega et al. \\
          & Aa-B       & 0.68      & 0.06                &                           &      2006 \\
2004.8232 & Aa-Ab      & 1.42      & 0.05                & 600/30                    &  This paper\\
          & Aa-B       & 1.27      & 0.05                &                           &              \\
\hline
\end{tabular}
\end{table}

We performed differential speckle photometry of the system with the 6 m
telescope of the Special Astrophysical Observatory of the Russian Academy of
Sciences and list the results in Table~\ref{tab2:Malogolovets_n}. We set the
$V$-band magnitude differences equal to $\Delta m_{AaAb}=1.35\pm0.06$ and
$\Delta m_{AaB}=1.11\pm0.06$, respectively, implying, given the corrected
parallax, the absolute magnitudes of  $M_{V}^{Aa}=7.31\pm0.08$,
$M_{V}^{Ab}=8.66\pm0.10$, and $M_{V}^{B}=8.42\pm0.10$, respectively. These
absolute magnitudes correspond to the spectral types of $Sp_{Aa}$ $\approx$K5,
$Sp_{Ab}$ $\approx$K9, and $Sp_{B}$ $\approx$K8, respectively. The above
spectral types agree well with the color index  $B-V$=1.24
(\cite{shtrass:Malogolovets_n}). The space velocity components
($U,V,W$)=(-75.5,-19.7,-42.3) (\cite{shtrass:Malogolovets_n}) and low emission
level in the  H and K Ca II lines (\cite{gray:Malogolovets_n}) imply that this
star should be classified as a Galactic-disk object with the age of 2--3 Gyr.

\section{ORBITAL PARAMETERS AND TOTAL MASSES}

The motion of components in a triple system can be subdivided
into two components: the motion about the center of mass of the
inner binary and the motion of the outer component about the
common center of mass. The orbit of the outer system AB was
computed by a number of authors from visual micrometric
measurements (\cite{baize:Malogolovets_n}; \cite{heintz:Malogolovets_n}).
Its main parameters are: a period $P$ of about 40 years; small
eccentricity $e$, and high inclination with respect to the sky
plane ($i$ $\approx$85$^{\circ}$). Soderhjelm
(\cite{soderh:Malogolovets_n}) refined the orbital elements of the
pair AB by combining the data of ground-based observations with
Hipparcos astrometry. He inferred a total mass of
$\Sigma\mathcal{M}_{AB}$=2.26$\pm$0.36$\mathcal{M}_{\odot}$,
which exceeds significantly the mass estimated by Duquennoy
(\cite{duq:Malogolovets_n}) based on spectroscopic and visual data.
Despite the use of Hipparcos astrometry, the outer orbit remains
uncertain, mostly because of its high inclination.

We determined the preliminary orbital parameters of the subsystems of  GJ\,795
with  allowance for new speckle interferomeric measurements based on the
Fourier transform of equations of motion (\cite{monet:Malogolovets_n}). At the
next stage, we refined the orbital elements via differential correction based
on the least squares method (see comments in \cite{forv:Malogolovets_n}). We
adopted early visual and several interferometric measurements from the
Washington Catalog of Binary Stars (\cite{wds:Malogolovets_n}). We set the
weights of speckle observations made with the 6 m telescope of the Special
Astrophysical Observatory of the Russian Academy of Sciences to be 10 times
higher than those of the visual and interferometric observations made with
other telescopes. The main reason why we attributed such low weight factors to
earlier data is that they did not allow for the binary nature of component A,
resulting in significant systematic errors.

\begin{figure*}[tbp]
\begin{center}
\includegraphics[scale=0.4]{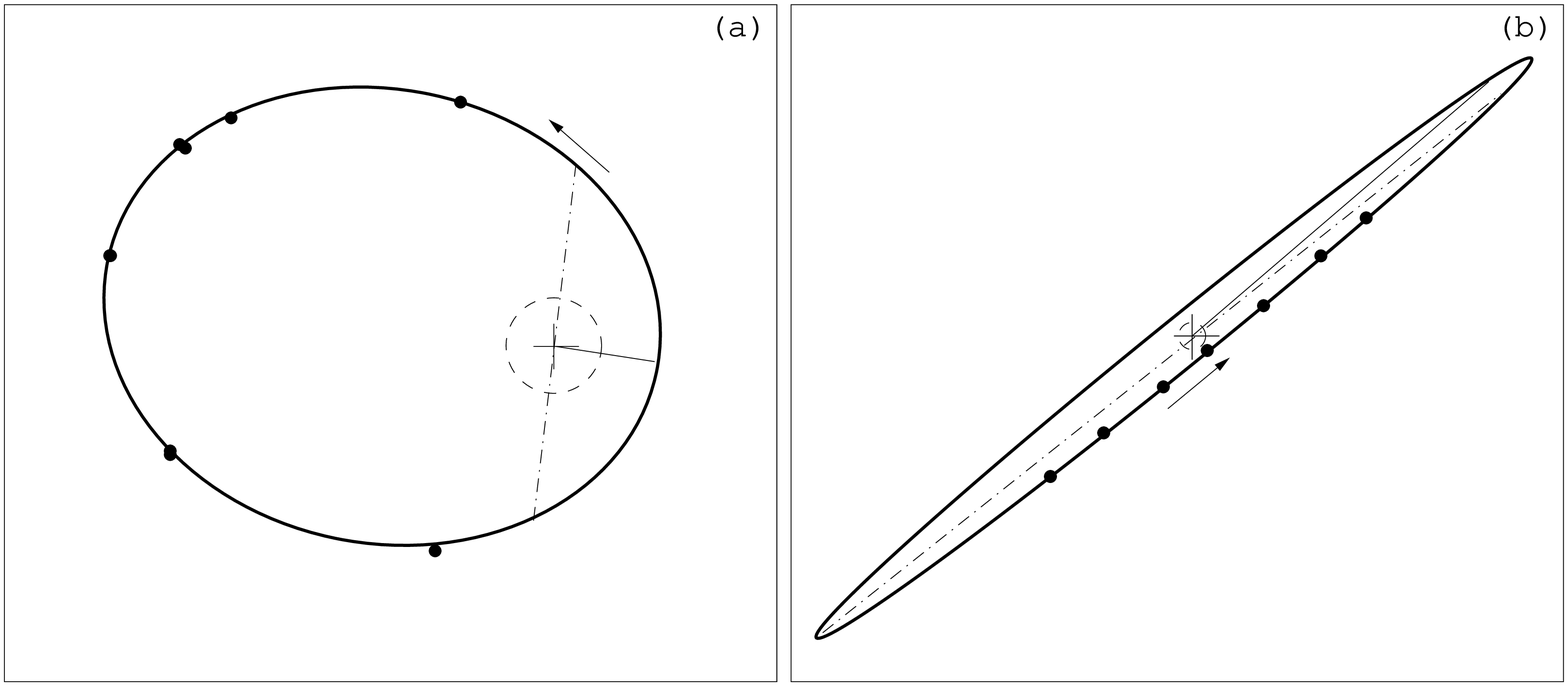}
\caption{Relative ellipses of the orbits of the triple
system GJ\,795: (a)
 orbit of binary Aab, (b)
 orbit of binary AB. The filled circles show the speckle interferometric
observations made with the 6 m telescope of the Special
Astrophysical Observatory of the Russian Academy of Sciences; the
solid line shows the position of the periastron, and the
dotted-and-dashed line, the line of nodes. The radius of the
dashed circle is equal to 20 mas. Position parameters of
component B are converted to the center-of-mass frame of binary
Aab.} \label{fig2:Malogolovets_n}
\end{center}
\end{figure*}

We computed the orbit of the inner binary Aab based exclusively
on the data of interferometric observations made with the 6 m
telescope of the Special Astrophysical Observatory of the Russian
Academy of Sciences. Seven measurements span over about 2.5
periods along the apparent ellipse of the orbit
(Fig.~\ref{fig2:Malogolovets_n}). An attempt to use radial
velocities from \cite{duq:Malogolovets_n} to construct a combined
orbit resulted in increased errors of the inferred orbital
elements. This is due to the combined effect of small radial
velocities of  Aab system, small inclination of the inner orbit
to the sky plane, and systematic errors in radial-velocity
measurements due to the influence of the distant component B.

To convert the motion of component B to the center-of-mass frame
of the inner binary Aab, we set the  mass ratio equal to
$q_{in}$=0.8 based on the empirical ``mass -- $M_{V}$'' relation
of \cite{henry:Malogolovets_n} and absolute magnitudes
$M_{V}^{Aa}$=7.3 and $M_{V}^{Ab}$=8.7. We give the elements of
the outer and inner orbits in Table~\ref{tab3:Malogolovets_n}. We
list all position measurements made with the 6 m telescope of the
Special Astrophysical Observatory of the Russian Academy of
Sciences and the corresponding residuals in
Table~\ref{tab4:Malogolovets_n}. The orbital parameters of the
inner binary  Aab agree well with those of the spectroscopic
orbit by Duquennoy (\cite{duq:Malogolovets_n}) and those of the
outer binary AB, with the refined orbit of Soderhelm
(\cite{soderh:Malogolovets_n}). It is remarkable that the
preliminary estimates of the inclination and semimajor axis of
the inner orbit obtained by Duquennoy (\cite{duq:Malogolovets_n})
based on published empirical ``mass -- luminosity'' relations
coincided exactly with their true values inferred from the
results of our interferometry.

\begin{table}
\begin{center}
\caption{Parameters of the inner and outer orbits in the triple system GJ\,795}
\label{tab3:Malogolovets_n}
\bigskip
\begin{tabular}{l|l|l}
\hline
           &        Aab         &  AB            \\
\hline
P, years           & 2.51$\pm$0.01     & 39.4$\pm$0.2    \\
T                  & 2000.55$\pm$0.01  & 1975.0$\pm$0.3  \\
e                  & 0.620$\pm$0.006   & 0.06$\pm$0.01   \\
a, mas             & 120$\pm$2         & 820$\pm$30      \\
i$\degr$           & 18$\pm$3          & 86.9$\pm$0.1    \\
$\Omega^{\circ}$   & 174$\pm$11        & 128.5$\pm$0.2   \\
$\omega^{\circ}$   & 87$\pm$11         & 212$\pm$2       \\
$\sigma_{\theta}$  & 0.5               & 0.8             \\
$\sigma_{\rho}$    & 1                 & 1               \\
\hline
\end{tabular}
\end{center}
\end{table}

\begin{table*}[p]
\begin{center}
\caption{Position parameters and residuals of the measurements of the triple
system GJ\,795} \label{tab4:Malogolovets_n}
\begin{tabular}{c|l|c|c|c|c}
\hline
Subsystem & Epoch & $\theta$ & $\rho$ & $(O-C)_{\theta}$ & $(O-C)_{\rho}$ \\
       &       &   degrees &    mas &       degrees &          mas \\
\hline
Aab& 1998.7741 & 55.0  &  161 &   0.3 & -2 \\
   & 1999.8206 & 105.7 &  164 &  -0.5 &  1 \\
   & 2000.8752 & 20.8  &  108 &   0.4 &  1 \\
   & 2001.7522 & 78.2  &  185 &  -0.2 &  0 \\
   & 2001.7522 & 78.5  &  186 &   0.5 &  1 \\
   & 2002.7986 & 150.3 &  98 &   0.9 &  2 \\
   & 2003.9272 & 61.9  &  172 &  -0.3 & -1 \\
   & 2003.9272 & 61.9  &  174 &  -0.3 &  1 \\
   & 2003.9272 & 61.7  &  174 &  -0.5 &  1 \\
   & 2004.8232 & 105.2 &  164 &  -0.1 &  0 \\
\hline
AB & 1998.7741 & 135.4 & 358 & 0.2  & 2  \\
   & 1999.8206 & 139.1 & 234 & -0.2 & 1  \\
   & 2000.8752 & 153.2 & 108 & 0.0  & 1  \\
   & 2001.7522 & 238.5 & 48 & 2.3  & 1  \\
   & 2001.7522 & 236.8 & 45 & 0.6  & -2 \\
   & 2002.7986 & 292.5 & 153 & 0.6  & 0  \\
   & 2003.9272 & 300.3 & 286 & 0.1  & -3 \\
   & 2003.9272 & 300.0 & 289 & -0.2 & 1  \\
   & 2003.9272 & 299.8 & 289 & -0.4 & 1  \\
   & 2004.8232 & 303.1 & 390 & 0.3  & -1 \\
\hline
\end{tabular}
\end{center}
\end{table*}

Let us now determine the angle  $\phi$ between the orbital planes:
\begin{equation}
\cos\phi=\cos i_{out}\cos i_{in}+
\nonumber\\
\sin i_{out}\sin i_{in} \cos(\Omega_{out}-\Omega_{in}),
\end{equation}
where $i_{out}$ and $i_{in}$ are the tilt angles of the orbit of
the outer and inner binaries with respect to the sky plane,
respectively; $\Omega_{out}$ and $\Omega_{in}$ are the longitudes
of the ascending node of the outer and inner binaries,
respectively. We now use the angle values from
Table~\ref{tab3:Malogolovets_n} to obtain $\phi$=74$^{\circ}$.

The dynamic mass of the inner binaries as inferred from the
orbital parameters and the corrected Hipparcos parallax of
$\pi_{Hip}^{\ast}$=58.8$\pm$2.1 mas (\cite{soderh:Malogolovets_n})
is equal to
$\Sigma\mathcal{M}_{Aab}$=1.28$\pm$0.15$\mathcal{M}_{\odot}$. The
mass of the entire system GJ\,795 computed using the orbital
parameters of binary AB is equal to
$\sum\mathcal{M}_{AB}$=1.69$\pm$0.27$\mathcal{M}_{\odot}$.
According to Lang (\cite{lang:Malogolovets_n}), the individual
masses of stars in the system as inferred from their absolute
magnitudes are equal to:
$\mathcal{M}_{Aa}$=0.67$\mathcal{M}_{\odot}$,
$\mathcal{M}_{Ab}$=0.57$\mathcal{M}_{\odot}$,
$\mathcal{M}_{B}$=0.54$\mathcal{M}_{\odot}$. The above estimates
imply a mass of
$\Sigma\mathcal{M}_{Aab}$=1.24$\mathcal{M}_{\odot}$ for Aab binary
and a total mass of
$\sum\mathcal{M}_{AB}$=1.78$\mathcal{M}_{\odot}$ for the entire
system, which are consistent within the quoted errors with the
total masses inferred using orbital parameters.

\section{HIERARCHY OF ORBITS AND THE KOZAI MECHANISM OF OSCILLATIONS}

The ratio of the orbital periods of the components of the system is equal to
$P_{out}$/$P_{in}$=15.7, and hence the system is weakly hierarchical and its
components form a single gravitationally bound system. Let us now try to
estimate its dynamical stability using empirical criteria and criteria based on
numerical simulations. According to one of such criteria suggested by Tokovinin
(\cite{tok2:Malogolovets_n}), the system is stable if the following inequality is
satisfied:
\begin{equation}
T=\frac{P_{out}(1-e_{out})^{3}}{P_{in}}>T_{c},
\end{equation}
where $P_{out}$ and  $P_{in}$ are the orbital periods of the
outer and inner binaries, respectively; $e_{out}$, the
eccentricity of the orbit of the outer binary, and $T_{c}$ is the
critical instability value, which is equal to 5. The stability
parameter for GJ\,795 is equal to T$\approx$13 and hence the
system is stable.

Based on numerical simulations, Harrington (\cite{harr2:Malogolovets_n}) proposed
the following stability parameter for triple systems, which depends on the
ratios of the semimajor axes of the outer and inner orbits, $a_{out}$ and
$a_{in}$, and the eccentricity of the outer orbit:

\begin{equation}
F=\frac{a_{out}(1-e_{out})}{a_{in}}>F_{c}.
\end{equation}

Like in the case of empirical estimate, the value of this parameter, $F$=6.42,
exceeds the critical level of $F_{c}$=5.46. However, these criteria should be
used with much caution when applied to systems with orthogonal orbits.

The semimajor axes of the outer and inner binaries are equal to
$a_{out}\approx$14 AU and $a_{in}\approx$2 AU,
respectively. The apoastron distance of the inner binary is equal
to $\approx$3.2 AU. The eccentricity of the outer orbit is close
to zero and therefore components of the triple are never at
comparable distances from each other.

Theoretical studies on the dynamics of multiple systems show that
in the case of large angles between the orbital planes the inner
and outer binaries (\cite{kozai:Malogolovets_n}) exchange angular
momentum. This mechanism triggers periodic variations (Kozai
oscillations) of the eccentricity of the inner orbit, $e_{in}$,
and angle $\phi$ between the orbital planes. The quantity
\mbox{$(1-e_{in}^{2})\cos^{2}\phi=const$} remains constant in this
process. The following formula gives the period of Kozai cycle:

\begin{equation}
P_{kozai}\sim P^{2}_{out}/P_{in}(1-e_{out})^{3/2}.
\end{equation}

The period of Kozai oscillations for the triple star GJ\,795 is
equal to only 560 years. We may try to directly observe Kozai
oscillations of the orbital parameters of binary Aab  over several
years of interferometric observations.

\section{CONSLUSIONS}

We use speckle interferometric observations made in 1998--2004
with the 6 m telescope of the Special Astrophysical Observatory
of the Russian Academy of Sciences to compute accurate visual
orbits for the outer and inner binaries of the triple star
GJ\,795. This triple system belongs to the disk component of the
Galaxy and is 2--3 Gyr old. Differential photometry of the
components of this system made it possible to construct a
complete model of GJ\,795, which agrees well with modern
empirical and theoretical relations. The absolute magnitudes of
the components are equal to $M_{V}^{Aa}$=7.31$\pm$0.08,
$M_{V}^{Ab}$=8.66$\pm$0.10, and $M_{V}^{B}$=8.42$\pm$0.10 and they
correspond to the spectral types of $Sp_{Aa}$ $\approx$K5, $Sp_{Ab}$
$\approx$K9, and $Sp_{B}$ $\approx$K8, respectively.

The orbital periods are equal to 2.51 and 39.4 years for the
inner and outer binaries, respectively. The angle between the
planes of the inner and outer orbits is equal to
$\phi$=74$^{\circ}$. The total dynamical masses
$\Sigma\mathcal{M}_{Aab}$=1.28$\pm$0.15$\mathcal{M}_{\odot}$ and
$\sum\mathcal{M}_{AB}=1.69\pm$0.27$ \mathcal{M}_{\odot}$ are
consistent with the estimated spectral types of the components.

We use the available empirical and theoretical stability criteria
to conclude that GJ\,795 is a gravitationally bound stable
hierarchical system. For objects of this type, the Kozai mechanism
should be efficient, which causes oscillations of the orbital
eccentricities and the angle between the orbital planes. During
its lifetime, the triple star GJ\,795, whose Kozai period is
equal to $P_{kozai}$ $\approx$560 years, must have undergone
$\sim10^{6}$ such periodic perturbations.

\begin{acknowledgements}
We are grateful to the night assistants at the 6 m telescope of
the Special Astrophysical Observatory of the Russian Academy of
Sciences for supporting the efficient work on the program. This
work was supported by the Russian Foundation for Basic Research
(grant no.~04-02-17563).
\end{acknowledgements}

\end{document}